\documentclass[useAMS,usenatbib]{mn2e}

\title[Chemical composition of NGC~5694]
  {{NGC~5694: another foster son of the Galactic Halo}
  \thanks{Based on data obtained at the Very Large Telescope under the programs 073.D-0211 and 089.D-0094.}
  }
   
\author[Mucciarelli et al.]
  {
  A. Mucciarelli,$^1$ M. Bellazzini,$^2$ M. Catelan,$^{3,4}$ E. Dalessandro,$^1$  P. Amigo,$^{3,4,5}$ 
   \newauthor
   M. Correnti,$^6$ C. Cort\'es,$^7$ 
    V. D'Orazi$^{8,9}$
  \\
  $^1$ Dipartimento di Fisica \& Astronomia, Universit\`a 
  degli Studi di Bologna, Viale Berti Pichat, 6/2 - 40127, 
  Bologna, Italy
  \\
  $^2$ INAF-Osservatorio Astronomico di Bologna, Via Ranzani, 1 - 40127, 
  Bologna, Italy
  \\
  $^3$ 
  Pontificia Universidad Cat\'olica de Chile, Facultad de F\'isica, 
  Departamento de Astronom\'ia y Astrof\'isica, Av. Vicu\~na Mackena 4860, \\
  782-0436 Macul, Santiago, Chile
  \\
  $^4$ 
  The Milky Way Millennium Nucleus, Av.\ Vicu{\~n}a Mackenna 4860, 782-0436 Macul, Santiago, Chile
  \\
  $^5$
  Departamento de F\'{i}sica y Astronom\'{i}a, Universidad de Valpara\'{i}so, Av. Gran Breta\~na 1111, Playa Ancha, Valpara\'{i}so, Chile
  \\
  $^6$  
  INAF-Istituto di Astrofisica Spaziale e Fisica Cosmica, Via Gobetti 101,
  I-40129, Bologna, Italy
  \\
  $^7$
  Departamento de F\'isica, Facultad de Ciencias B\'asicas, Universidad Metropolitana de Ciencias 
  de la Educaci\'on, Av. Jos\'e Pedro \\ Alessandri 774, 
  776-0197 \~{N}u\~noa, Santiago, Chile
  \\
  $^8$ 
Department of Physics and Astronomy, Macquarie University, Balaclava Rd, North Ryde, NSW2109, Australia
\\
  $^9$
Monash Centre for Astrophysics, School of Mathematical Sciences, Building 28, Monash 
University, VIC3800, Australia 
\\
  }

\usepackage{graphicx}

\pagerange{\pageref{firstpage}--\pageref{lastpage}} \pubyear{2002}

\def\LaTeX{L\kern-.36em\raise.3ex\hbox{a}\kern-.15em
    T\kern-.1667em\lower.7ex\hbox{E}\kern-.125emX}

\begin{document}

\label{firstpage}

\maketitle

\begin{abstract}
We present the results of the analysis of high-resolution spectra obtained with UVES-FLAMES@VLT 
for six red giant branch stars in the outer-halo metal-poor ([Fe/H]I=-1.98 and [Fe/H]II=-1.83) Galactic globular cluster 
NGC~5694, which has been suggested as a possible incomer by Lee et al. (2006) based on the anomalous 
chemical composition of a single cluster giant. We obtain accurate abundances for a large number 
of elements and we find that: 
(a)the six target stars have the same chemical composition within the uncertainties, except for 
Na and Al; (b) the average cluster abundance of $\alpha$ elements (with the only exception of Si)
is nearly solar, at odds with typical halo stars and globular clusters of similar metallicity; 
(c) Y, Ba, La and Eu abundances are also significantly lower than in Galactic field stars and star 
clusters of similar metallicity. Hence we confirm the Lee et al. classification of NGC~5694 as a 
cluster of extra-galactic origin. We provide the first insight on the Na-O and Mg-Al anti-correlations 
in this cluster: all the considered stars have very similar abundance ratios for these elements, except 
one that has significantly lower [Na/Fe] and [Al/Fe] ratios, suggesting that some degree of early 
self-enrichment has occurred also in this cluster.
\end{abstract}


\section{Introduction}

Modern wide-area optical surveys have revealed that the stellar halo of nearby galaxies (including our own) 
contains a wealth of substructures due to the accretion of smaller haloes 
\citep[see, e.g.,][and references therein]{iba07,bell}, as expected in a hierarchical merging scenario for 
galaxy formation \citep[e.g.,][and references therein]{bj05,abadi06}. It is now observationally established 
that the accretion of satellites leads also to the accretion of their globular clusters 
\citep[GC;][]{bfi03,lm10,mackey10,mackey13}. Therefore, searching for Galactic GCs that can be identified as 
coming from disrupted satellites is one possible reasearch lines to investigate the assembly history 
of the Milky Way halo and the associated GC system.

A powerful technique to spot individual members of an ancient accretion event (either a star or a star cluster) 
consists in finding some peculiar feature in their elemental abundance pattern markedly different from typical 
Halo stars, the so-called {\em chemical tagging} 
\citep[see][for recent applications, discussion and references]{free,schla_tag,mitsh_tag}. 
Examples of the identification of likely accreted GCs by means of chemical tagging are provided by \citet{cohen04} 
for Palomar~12, and by \citet{mottini} for Ruprecht 106. 

In this framework, \citet[][L06 hereafter]{lee} obtained high-resolution spectroscopy of a bright red giant branch 
(RGB) star in the Galactic cluster NGC~5694, already indicated as a possible incomer by \citet{HH76} because of 
its large distance and velocity 
\citep[$R_{GC}\simeq 30$~kpc, $V_r\simeq -140$~km/s;][]{correnti,geisler95}.
Intriguingly, \citet{correnti} found that the cluster density profile is much more extended 
than previously believed and declines, in the outer fringes of the cluster, with a slope that is not 
reproduced by model profiles generally adopted for classical GCs.
L06 derived the chemical abundance for several elements in the considered star and revealed that it displays 
[$\alpha$/Fe], [Ba/Fe] and (possibly) [Cu/Fe] ratios significantly different from those of typical halo stars and 
of the bulk of Galactic GCs of similar metallicity, concluding that the cluster has an extra-galactic origin. 
This conclusion is clearly very interesting, but it is based on a single cluster star, virtually located at 
the very tip of the RGB. 

To obtain a clear-cut confirmation of the tagging of NGC~5694 proposed by L06 we obtained high-resolution 
spectra of {\em six} cluster members.  
In this paper we present the results of the analysis of these spectra. We obtained elemental abundances 
that fully confirm (and extend) the results by L06. In addition, we provide the first insight into the early 
self-enrichment history of this cluster, as traced by the spread in Na, O, Mg, and Al, that has been recently 
recognized as a common feature of old GCs \citep[see][and references therein]{gratton}.

\section{Observations}
\label{obs}

\begin{figure}
\includegraphics[width=84mm]{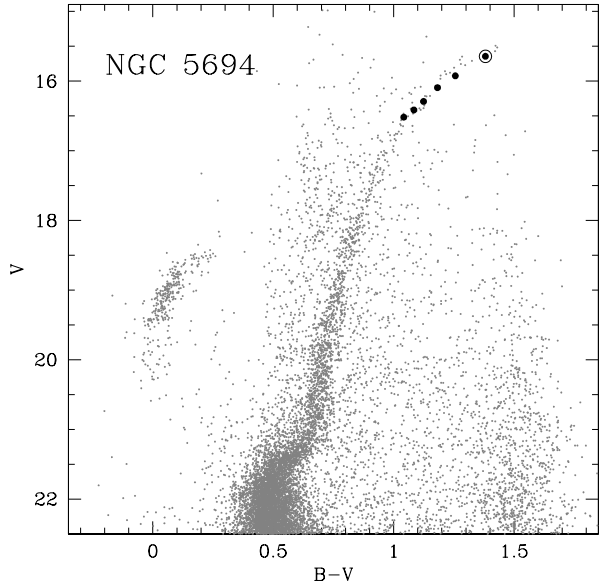}
\caption{CMD of NGC~5694: the stars analyzed in this paper are highlighted as filled circles 
(the empty circle marks the star 
\#37, in common with L06).}
\label{cmd}
\end{figure}

The data have been acquired with the FLAMES@VLT spectrograph in the combined MEDUSA+UVES mode, 
allowing the simultaneous allocation of 8 UVES high-resolution fibers (discussed in this paper) 
and 132 MEDUSA mid-resolution fibers (discussed in a forthcoming paper, mainly devoted to the internal kinematics of the cluster). 
The employed spectral configuration for the UVES targets discussed in this paper is 
the 580 Red Arm, with a spectral resolution of $\sim$40000 and covering the range $\sim$4800--6800 $\mathring{A}$.
A total of 8 exposures of 46 min each for the same targets configuration has been secured in Service Mode 
during the period between April and July 2012. The targets have been selected in the bright portion 
of the RGB (V$<$16.5) from the B,V photometric catalog by 
\citet[][C11 hereafter; from VIMOS@VLT observations over a $24\prime\times20\prime$ field of view]{correnti}, 
complemented with publicly available WFPC2@HST photometry of the central region from \citet{snap}. 
We selected our UVES targets
among stars already confirmed as cluster members from their radial velocity by \citet{geisler95}; among them, 
we included also the only star  analyzed by L06, namely our star \#37. 
To avoid significant contamination from other sources, we excluded from the original target list 
any star (of magnitude V$_0$) having a companion with V$<$V$_0+$1.0 within two arcsec of its center. 
Two out of the eight available UVES fibers have been used to sample the sky background and perform a robust 
sky subtraction for each individual exposure. The position of the
six target stars in the cluster Color Magnitude Diagram (CMD)  is shown in Fig.\ref{cmd}, and their 
basic parameters are reported in Table~1.

The data reduction has been performed using the last version of the FLAMES-UVES 
CPL-based ESO pipeline\footnote{http://www.eso.org/sci//software/pipelines/}, including 
bias-subtraction, flat-fielding, wavelength calibration with a standard Th-Ar lamp and 
spectral extraction. The dispersion solution has been checked by measuring the position 
of several sky emission lines and comparing them with their rest-frame position taken from 
the sky lines atlas by \citet{oster96}, finding no relevant wavelength shifts. We found that 
the final merged spectra obtained from the ESO pipeline are not satisfactory because of the quite poor 
quality in the overlapping region of adjacent orders. Thus, we decided to perform the order merging 
under IRAF.

Each single (sky-subtracted) exposure of a given star has been converted to the heliocentric system (the 
heliocentric correction has been computed with the IRAF task {\sl rvcorrect}) and finally 
coadded together in order to obtain a median spectrum (thus removing spikes, cosmic rays and other 
random spectral impurities). The typical SNR per pixel for the brightest star are $\sim$60 at 
$\sim$5300 $\mathring{A}$ and $\sim$80 at $\sim$6300 $\mathring{A}$; for the faintest star the SNR are $\sim$30 and 
$\sim$50, at $\sim$5300 and $\sim$6300 $\mathring{A}$, respectively.

\subsection{Radial velocities}

Radial velocities (RV) have been derived using the DAOSPEC code \citep{stetson}, through the 
identification of $\sim$250-300 absorption lines in each star. The two UVES chips are analyzed independently; 
the agreement between the two radial velocity estimates is excellent
($<RV^{lower}$-$RV^{upper}>$=--0.28 km/s, $\sigma$=~0.24 km/s).
The typical uncertainty arising from this procedure (internal error) is $\sim$0.05 km/s, but the dominant 
source of uncertainty is the zero-point of the wavelength calibration that we estimated by measuring the position 
of several emission and absorption telluric features. The uncertainty in the wavelength calibration (typically 0.3-0.5 km/s) 
is added in quadrature to the internal error. 
The heliocentric RV and total uncertainty for each target star are listed in Table~1.

The average RV is $V_r=-140.4\pm2.2$ km/s ($\sigma$=5.4 km/s). 
This value nicely  agrees with those derived by \citet{geisler95} and L06.
The mean difference between our RV and those listed by \citet{geisler95} is of 
$<RV_{\rm this~work}$-$RV_{\rm Geisler~et~al.(1995)}>$=+0.6 km/s.
with a dispersion of $\sigma=6.9$~km/s, that can be considered as satisfying given the 
low-resolution of the spectra analyzed by \citet[][$R=\simeq$ 3000]{geisler95}.
Also, we find an excellent agreement ($<RV_{\rm this~work}$-$RV_{\rm L06}>$=+0.0 km/s) 
with the RV value provided by L06
for the star \#37, by using spectra with a spectral resolution similar to that of UVES.

\section{Chemical analysis}

The determination of the chemical abundances for the majority of the elements 
has been obtained through the measurement of the equivalent widths (EWs) of atomic transition lines 
and employing the package GALA \citep{m_gala} based on the WIDTH9 code by R. L. Kurucz. 
For the lines that cannot be treated with the EWs method (because they are blended 
with other lines) we resort to the comparison between the observed and synthetic spectra.
Model atmospheres have been computed with the last version of the ATLAS9 code, 
assuming local thermodynamic equilibrium (LTE) for all the species and one-dimensional, plane-parallel 
geometry. In the flux computation we cannot take into account the {\sl approximate} overshooting 
\citep{castelli97}. 
The opacity distribution functions employed in the model calculations are those available 
in the website of F.Castelli \footnote{http://wwwuser.oat.ts.astro.it/castelli/odfnew.html}  
computed for a metallicity [M/H]=--2.0 dex 
(according to L06) and with solar-scaled abundance patterns for 
each element of the chemical mixture. 

All the chemical abundances have been derived under the assumption of LTE.
Corrections for departures from the LTE approximation are applied only for the Na lines, by  
interpolating on the grid of corrections calculated by \citet{lind}. The grid by \citet{lind} 
does not include the cases with log~g$<$1, while three program stars have surface gravity slightly 
lower than this boundary. For these stars the corrections have been computed assuming log~g=~1. 
Since the corrections are small and only weakly dependent on stellar parameters - in the considered range - 
this approximation should have a negligible impact on the final abundance estimates.

Reference solar abundances are from \citet{gs98} except for oxygen, for which we adopted the most 
recent value by \citet{caffau}.

\subsection{Atmospheric parameters}
Due to the combination of low metallicity and low surface gravity of our targets, a fully 
spectroscopic derivation of all the atmospheric parameters could be biased from departures 
from LTE conditions (as discussed in Section~\ref{nlte}). 
In order to avoid this kind of problems, we resort to photometry to derive effective temperatures
($T_{eff}$) and surface gravities (log~g). Atmospheric parameters have been derived from B,V photometry in the C11 catalog (with the expansion described in Sect.~\ref{obs}). 

$T_{eff}$ values have been computed by means of the $(B-V)_0$--$T_{eff}$ 
transformation by \citet{alonso99} based on the Infrared Flux Method; the dereddened color 
$(B-V)_0$ is obtained adopting a color excess E(B-V)=~0.099 mag (C11) and the 
extinction law by \citet{mccall}. 
Surface gravities have been computed with the Stefan-Boltzmann relation, assuming the 
photometric $T_{eff}$, the distance modulus of 17.75$\pm$0.10 (C11) and an evolutionary mass 
of 0.75 $M_{\odot}$, as estimated from the isochrone from the BaSTI dataset \citep{pietrinferni04} 
with age of 12 Gyr, Z=~0.0003 and a solar-scaled chemical mixture (according to L06)
\footnote{The choice of the stellar mass is affected by the details of the mass loss, a process not easy 
to model and still poorly known. However, the adoption of different values of the stellar mass has a 
very small ($<$0.02-0.03 dex) impact on the derived abundances.}.
The bolometric corrections are calculated according to Eq.~(17) of \citet{alonso99}.
Microturbulent velocities ($v_t$) have been estimated by minimizing the slope between the neutral 
iron lines and their reduced EWs (obtained as the logarithm of the EW normalized to 
its wavelength).

Uncertainties in $T_{eff}$ are estimated by taking into account the uncertainty in B and V 
magnitudes and in the color excess.
For the very bright stars considered here, the dominant factor in the photometric error 
is the uncertainty in photometric zero-point: here we conservatively assume a photometric error of 
0.03 mag in both filters for all the targets. For the color excess E(B-V) we adopted an uncertainty of 0.02~mag. 
The total uncertainty (photometry + extinction) 
corresponds to a typical uncertainty in $T_{eff}$ of $\sim$50-60 K.

Uncertainties in the surface 
gravity are derived by considering the error sources in $T_{eff}$, 
luminosity (including the uncertainties of the photometry and in the distance modulus) and in mass 
(that we assume $=\pm 0.05 M_{\odot}$, corresponding to the typical spread in total mass 
along the horizontal branch of globular clusters, see e.g. \citet{val08} ). 
Summing all these terms in quadrature leads to a typical uncertainty in log~g of 0.06~dex.

Finally, the error associated to the determination of $v_t$ is estimated by propagating the 
uncertainty in the slope in the reduced EW vs Fe abundance plane, through a Jackknife bootstrap 
technique \citep{lupton}. This uncertainty is as small as $\sim$0.08 km/s, thanks to the 
large number of Fe~I lines used to determine $v_t$.
To this value we added in quadrature the uncertainty in $v_t$ due to the 
covariance with the other parameters: $T_{eff}$ has been varied by $\pm$1$\sigma_{T_{eff}}$ 
(thus leading to a variation of log~g)
and the procedure to derive $v_t$ has been repeated. Finally, we estimated a typical 
error in $v_t$ of $\pm$0.15 km/s.


\subsection{A comment about LTE and NLTE}
\label{nlte}
Late-type stars with low metallicities and surface gravities (i.e. approaching 
the RGB-Tip) are expected to suffer from deviations from LTE conditions due to the occurrence 
of over-ionization because of the UV radiation excess 
\citep[basically $J_{\nu}>B_{\nu}$, where $J_{\nu}$ is the source function and $B_{\nu}$ 
is the Planck function, see e.g.][]{asplund}.

The occurrence of these NLTE effects thwarts the derivation of log~g from the ionization equilibrium, 
i.e. from the requirement that Fe~I and Fe~II are the same within the uncertainties. Basically, a forced 
ionization equilibrium in over-ionization conditions leads to underestimate the surface gravities.
For our targets we found, on average, that $<$Fe~I - Fe~II$>$ = --0.13 dex ($\sigma$=~0.05 dex). To erase 
this offset we need to adopt lower log~g values, typically between 0.0 and 0.3, incompatible with 
the stellar evolution predictions for a low-mass star in the bright portion of the RGB, in an old and metal-poor 
stellar system as NGC~5694. For instance, the isochrone used to derive the gravities predicts 
a surface gravity of 0.47 for its brightest point. Note that a fully spectroscopic analysis leads to 
temperatures lower by $\sim$100 K than those predicted from the isochrone.  
Also, we checked that this offset cannot be a spurious effect due to the adopted log~gf values: 
an analysis performed on the solar flux spectrum of \citet{neckel}, adopting the solar ATLAS9 model atmosphere 
computed by F. Castelli\footnote{http://wwwuser.oat.ts.astro.it/castelli/sun/ap00t5777g44377k1odfnew.dat}, 
provides Fe~I-Fe~II=--0.01$\pm$0.02 dex, excluding systematic offsets between the oscillator 
strengths of Fe~I and Fe~II lines.
Thus, we can easily attribute the discrepancy between Fe~I and Fe~II abundances in the stars of NGC~5694 to 
NLTE effects. 


In order to reduce the effects due to over-ionization, we decide to calculate the abundance 
ratios from neutral lines by scaling the abundance on the Fe~I abundance and those from single ionized lines 
by scaling the abundance on Fe~II. In fact, lines of the same ionization stage have similar dependencies 
from the parameter (in particular from the gravity).
Among the elements derived from neutral lines, only the [O/Fe] abundance ratio has been scaled on Fe~II, 
because the over-ionization effects are reduced from its first ionization potential and the oxygen abundance 
(when derived from the forbidden O line at 6300.3 $\mathring{A}$, as we have done)
has a sensitivity to the gravity similar to that of the ionized elements
\citep[we refer the reader to the detailed discussion in][]{ivans01}.

\subsection{Line list}

We selected the transitions to be analyzed through the visual inspection of synthetic spectra 
calculated with the atmospheric parameters of the targets (and assuming a global metallicity 
[M/H]=--2.0 dex, with solar-scaled abundance ratios for all the elements) 
and convolved with a Gaussian profile at the UVES resolution. 
The selected lines are chosen unblended from other transitions. 
Oscillator strengths and excitational potential are taken from the last version of the Kurucz/Castelli 
line-list\footnote{http://wwwuser.oat.ts.astro.it/castelli/line-lists.html}. 
When available, the Van der Waals damping constants for the broadening by neutral hydrogen collisions  
calculated by \citet{barklem00} have been adopted, and for 
the other transitions we adopted the values from the Kurucz database or, when the Van der Waals constants are 
not available in the literature, they are computed by adopting the approximate formulas discussed 
in \citet{castelli_wid}. 
After a first analysis, we refine the line-list repeating the procedure of the lines selection by using 
synthetic spectra computed with the proper chemical composition of each star; typically, only a few 
lines are discarded at this step because they are blended with other transitions, and 
a few new transitions added to the line-list. Finally, the analysis is repeated with the clean line-list.

\subsection{Abundances from equivalent widths}
\label{ew}

EWs have been measured with the code DAOSPEC \citep{stetson} that 
automatically performs the line profile fitting adopting a saturated Gaussian function, defined as 
h($\lambda$)=$\frac{g(\lambda)}{1+g(\lambda)}$, where $g(\lambda)$ is the value of the Gaussian function 
at a given wavelength $\lambda$. All the spectral lines are fitted by adopting a fixed Full Width at Half
Maximum (FWHM), that scales with the wavelength, as in wavelength-calibrated echelle spectra 
the spectral resolution remains constant along the entire spectrum and the FWHM varies with
$\lambda$.

The FWHM for each spectrum has been estimated iteratively: we performed a first analysis using as 
input value the nominal FWHM of the grating, leaving DAOSPEC free to re-adjust the FWHM  in order 
to minimize the median value of the residuals distribution. The new value of the FWHM (referred to the 
center of the spectrum) is used as input value for a new run, until a convergence of $\sim$0.1 pixel 
is reached. 
After a visual inspection of the spectra and the check of the final residuals, we adopted for the 
normalization a $3^{\rm rd}$-order Legendre polynomial. We checked that the use of a slightly different 
degree changes the derived FWHM at a level of 0.1-0.2 pixels, with a negligible impact on the EWs.

We include in the chemical analysis only lines that satisfy the following criteria: 
\begin{itemize}

\item EW error smaller than 20\%. As EW uncertainty we assumed the error 
estimated by DAOSPEC from the standard deviation of the local flux residuals and 
representing a 68\% confidence interval of the derived EW. An error in EW of $\pm$20\% translates 
into an abundance error of $\pm$0.1 dex for a line located in the linear part of the curve of growth 
(typically, the lines with EW errors larger than $\sim$10\% are the weakest ones).
The median abundance error due to the EW uncertainty for our lines is of $\pm$0.05 dex (only a few lines 
have abundance errors larger than 0.1 dex);

\item a reduced EW smaller than --4.7, corresponding 
to $\sim$100 m$\mathring{A}$ for a line at 5300 $\mathring{A}$ and $\sim$120 m$\mathring{A}$ for a line at 6300 $\mathring{A}$. This 
boundary allows to exclude lines in the flat part of the curve of growth, that are less sensitive to 
the abundance, highly sensitive to the microturbulent velocity and the velocity fields, and for which 
the Gaussian approximation starts to fail, because of the development of Lorentzian wings in the line profile;

\item a reduced EW larger than --6, corresponding to $\sim$5 and $\sim$6 m$\mathring{A}$ at 
5300 and 6300 $\mathring{A}$, respectively. This boundary excludes lines that are too weak and noisy;

\item DAOSPEC quality parameter Q lower than 1.2 in all the spectra. This 
parameter is the ratio between the local intensity residual of a given line and the 
root-mean-square intensity residual of the entire spectrum. 
Typically, only a couple of lines have been discarded because their Q parameters are systematically 
higher than the adopted threshold in each spectrum.
\end{itemize}

At the end of the analysis, the mean abundance (and the corresponding dispersion) for a given element in a given star is 
computed by averaging the abundances of the surviving lines weighted by the uncertainty on the abundance as obtained 
from the EW error. 
For those element for which only one line is available, 
the error associated to the individual abundance estimate is taken as the internal error.

Table~2 lists some measured EWs and atomic data, the complete set being available in the 
online journal.

\subsection{Abundances from spectral synthesis}
\label{synt}

We compare the observed spectra with synthetic spectra to derive the abundance 
from those lines that exhibit a composite line structure. 
The oxygen abundance has been derived from the 
forbidden line at 6300.31 $\mathring{A}$\footnote{This transition is often superimposed on telluric lines 
both in absorption and emission; 
we carefully checked that in all the targets the oxygen line is not contaminated by telluric features.}, 
including the contribution of a close Ni line (for each star we adopted the corresponding Ni abundance 
derived from EWs).
Odd-Z, iron-peak elements (Sc~II, V~I, Mn~I, Co~I and Cu~I) suffer from hyperfine splitting and 
we adopted for those transitions the line-list provided in the Kurucz 
database\footnote{http://kurucz.harvard.edu/line-lists/gfhyper100/}.
For Cu the only available transition is that at 5105 $\mathring{A}$, because the other Cu line usually investigated (at 5782 $\mathring{A}$) 
falls in the gap between the two UVES chips.
For Ba~II we measured the red lines at 5853.7, 6141.7 and 6496.9 $\mathring{A}$ using the line-list available in the 
NIST database taking into account both hyperfine structure and isotopic splitting. 
Other Ba~II lines are available in the blue part of our spectra but we excluded these transitions 
because they can be affected by relevant NLTE effects \citep[see][]{sneden96}.
We derived only 3$\sigma$ detection upper limits 
for the strongest La~II and Eu~II available lines, at 
6390.5 and 6645.1 $\mathring{A}$, respectively. Also for these two lines, the modeling of synthetic spectra 
includes the hyperfine structure and (only for Eu~II) isotopic splitting, by using the line-list 
by \citet{lawler01a} and \citet{lawler01b} for La and Eu, respectively.

The abundance for each selected line has been derived through a $\chi^2$-minimization between 
the observed spectrum and a grid of synthetic spectra calculated at different abundances. 
The synthetic spectra have been computed by means of SYNTHE code developed by R. L. Kurucz, 
including the entire Kurucz/Castelli line-list (both for atomic and molecular transitions) and 
adopting the same model atmospheres used to derive the abundances from EWs.
The synthetic spectra are convolved at the instrumental resolution, 
then rebinned at the same pixel-step of the observed spectra (0.014 and 0.017 $\mathring{A}$/pixel for 
the lower and upper chip of the 580 Red Arm grating, respectively). Additional details about 
the fitting procedure are discussed in \citet{m12}.

The uncertainty in the line fitting procedure has been estimated by means of Monte Carlo simulations:
for each line the fitting procedure has been repeated by using a sample of 500 synthetic spectra where Poissonian 
noise has been injected (after the re-mapping of the synthetic spectra at the same pixel-step of the UVES spectra)
in order to reproduce the noise conditions observed around the analysed line. These uncertainties (ranging from 
$\sim$0.02 dex up to $\sim$0.07 dex, for SNR=~70 and 30, respectively for a line of moderate strength) 
have been used as weights to compute the average abundance for a given species or 
as internal error when only one line is available, as done for abundance estimates derived from EWs.

\begin{figure}
\includegraphics[width=84mm]{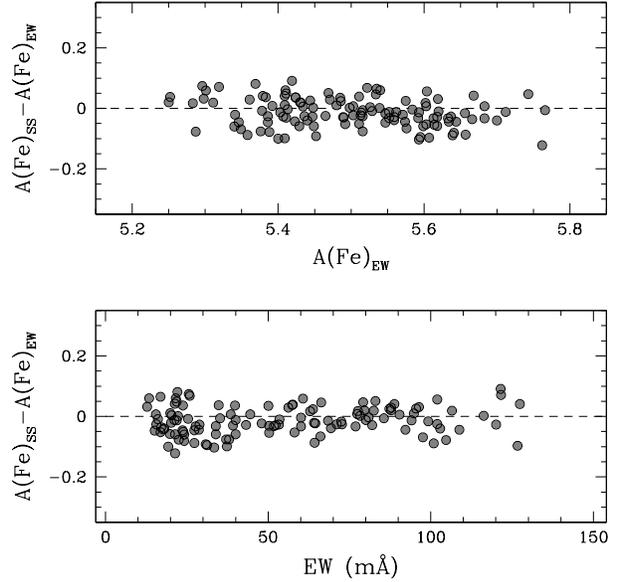}
\caption{ Difference of the abundances of neutral iron lines for the star \#37 as inferred from spectral synthesis and 
EW measurements as a function of the iron abundance (from EWs, upper panel) and 
of the EW (lower panel). Horizontal dashed lines mark the zero value.}
\label{testss}
\end{figure}

In order to check possible systematic offsets between the abundances inferred with the 
two methodologies (EWs and spectral synthesis) we re-analysed with the method described above 
the Fe~I lines for the star \#37 (taken as representative of our sample). 
Fig.~\ref{testss} shows the difference between the A(Fe~I) abundances inferred with the 
two methods as a function of the iron abundances derived from EWs (upper panel) and 
of the EWs (lower panel). The average difference turns out to be $A(Fe~I)_{SS}-A(Fe~I)_{EW}$=~-0.013 dex 
($\sigma$=~0.047 dex); also, none specific behaviour with the line strength is found. This check 
confirms that the two methods provide basically the same results.

\subsection{Abundance uncertainties}
\label{unc}
The total uncertainty for a given elemental abundance is computed as the sum in quadrature 
of two sources of error: (1) that arising from the measurement procedure (EW or spectral 
synthesis), and (2) that arising from the atmospheric parameters. 
The measurement errors are computed as described in Sections~\ref{ew} and \ref{synt}.

Internal uncertainties related to the measurement errors for each elemental abundance 
are computed as the dispersion of the mean normalized to the root mean square of the 
number of used lines.
Uncertainties due to atmospheric parameters are calculated by varying each time only one parameter 
by the corresponding error, keeping the other ones fixed and repeating the analysis.

As discussed by several authors \citep[see e.g.][]{mcwilliam95,barklem05}, when we obtain 
the abundance ratios (as [X/Y[=[X/H]-[Y/H]) the systematic uncertainties related to the atmospheric parameters 
partially cancel out, because, in general, lines of the same ionization stage respond in a similar 
way to changes in the stellar parameters. 
For instance, the estimated abundance variation due to 
$\pm1\sigma_{T_{eff}}$ for the star \#37 is of $^{+0.07}_{-0.06}$ for Fe and of 
$^{+0.07}_{-0.07}$ for Ca. Therefore, an error in $T_{eff}$ of that amplitude translates into variations of the 
two considered quantities amounting to just $^{+0.00}_{-0.01}$ in the corresponding [Ca/Fe] abundance ratio. 
On the other hand, the internal errors due to 
the (EWs or spectral synthesis) measurements are independent and they will be 
combined in quadrature for each abundance ratio. Thus, we provide for each element two different 
estimates of the total uncertainty: (1)~the uncertainty for the absolute abundance [X/H] obtained 
by combining in quadrature the internal error and those related to each atmospheric parameter, and 
(2)~the uncertainty for the abundance ratio [X/Fe] (where the reference iron abundance is that 
from Fe~I lines for neutral elements and from Fe~II lines for single ionized elements) obtained 
by summing in quadrature the internal errors of X and Fe, and the relative abundance variation 
due to each atmospheric parameter.

\subsection{Results and comparison with L06}

Table~3 summarizes the measured abundance ratios in the target stars, together with the adopted 
reference solar value and the total uncertainty.  Reported errors are the total uncertainties
of the [X/Fe] abundance ratios, while the values in the brackets are the uncertainties of the [X/H] abundance ratios (see Sect.~\ref{unc}).

The inclusion in our sample of the star studied by L06 (our star \#37) give us the opportunity to check the reliability of our abundances by the comparison with a completely independent analysis, based on independent photometry and high-resolution spectra (from MIKE@MAGELLAN). The elemental abundances for Fe, Mg, Si, Ca, Ti, Mn, Ni, Cu, Y, Ba, La and Eu have been obtained in both studies, for this star.

The atmospheric parameters adopted in the two analyses are very similar ($\delta T_{eff}$=~20 K, $\delta$log~g=~0.05 and $\delta v_{t}$=~0.3 km/s).
We find small differences in the iron content, with $Fe_{us}-Fe_{Lee06}$=+0.07~dex and 
+0.10~dex for neutral and single ionized Fe lines.
Basically, these differences are fully ascribable to the differences in the atmospheric parameters 
(mainly the difference in $v_t$).
We re-analyse this star by adopting the same atmospheric parameters as in L06. 
Considering also the small offset in the employed solar iron abundance
($\delta Fe_{\odot}$=~0.02), the difference turns out to be of +0.02 for [Fe/H]~I and 
0.0 for [Fe/H]~II. 

For the other abundance ratios, the typical offset are smaller than 0.1 dex, except for 
Mn (${\rm [Mn/Fe]}_{us}-{\rm [Mn/Fe]}_{Lee06}$=--0.22 dex) and 
Ba (${\rm [Ba/Fe]}_{us}-{\rm [Ba/Fe]}_{Lee06}$=--0.24 dex). However also these differences are  within
the combined uncertainties of the two estimates.  
At least for Ba, the offset 
can be easily ascribed to the different $v_t$ values adopted 
in the two analyses, because the Ba~II lines are located in the flat part of the curve of growth 
and they are sensitive to the velocity fields.
Unfortunately, L06 do not provide details about the employed hyperfine structure, 
oscillator strength and isotopic splitting for the lines derived by spectral synthesis and 
we are thus unable to properly compare the corresponding abundances.

For La and Eu, the spectrum acquired by L06 has a higher SNR with respect to our one, thus 
they are able to properly measure the abundances for these elements, while we can provide only 
upper limits. However, our upper limits are consistent with their results.

It may be worth noting that star \#37 has been recognized as a semi-regular variable by 
\citet[][their V12]{marcioRR}. The V magnitude of the star in the 26 epochs sampled by these authors, 
spread over $\sim 100$~days, shows a typical r.m.s. scatter of a few hundredths of a magnitude 
(a maximum $\Delta V\la 0.1$~mag). This suggests that the amplitude of the variation is so small as have 
negligible effect on the chemical abundance analysis, as confirmed also by the excellent agreement between 
our results and those by L06, derived from data obtained in different epochs and as described above.

\subsection{Defining a test case for differential comparison: NGC~6397}
\label{6397}

Since the main scientific goal of the present study is to verify if NGC~5694 has indeed a chemical abundance pattern 
setting it apart from the bulk of classical Galactic GCs of similar metallicity, it seems safe to test the hypothesis 
against a bona-fide, {\em typical} 
MW halo cluster, by adopting {\em a strictly homogeneous analysis on very similar data}.

With this aim, we retrieved, from the ESO Archive\footnote{http://archive.eso.org/cms/}, the UVES@FLAMES red-arm spectra of 13 giant 
stars of the GC NGC~6397 \citep[that has the same metallicity of NGC~5694 and an abundance pattern typical of the MW halo, see e.g.][]{carretta09,lind10,lovisi12}
taken with the same instrumental configuration used here for NGC~5694.

Atmospheric parameters have been determined with the same technique described in Section 4.1, 
adopting the B, V photometry used by \citet{carretta09} and following the same steps of the 
analysis described above.
Even if these giants of NGC~6397 are slightly less evolved with respect to those of NGC~5694 
(with $T_{eff}$ in the range $\sim$4700-4900 K and log~g between 1.5 and 2.0), we analysed them 
with the line-list optimized for the targets of NGC~5694, to maintain the full homogeneity of the analysis. 
For this reason, some transitions detectable in our NGC~5694 stars cannot be measured in the stars of the 
NGC~6397 sample because they are too weak in their warmer atmospheres (in spite of the nearly identical metallicity 
of the two clusters). The derived average abundances for a number of elements are reported in Table~4
(together with the differences $\delta$ between the abundance ratios measured in NGC~5694 and NGC~6397) and will 
be discussed in the following, providing an extremely robust differential benchmark for the comparison with NGC~5694 
(see Sect.~\ref{homo} for further details on the content of this table).
 
Here we simply note that from our analysis, NGC~6397 has [Fe/H]I=--2.00$\pm$0.02 and [Fe/H]II=~-1.90$\pm$0.01, in nice agreement 
with previous studies \citep[][and references therein]{lind10,koch11,lovisi12}. 
In particular, with respect to the 
analysis by \citet{carretta} based on the same spectra, we find an average difference 
${\rm [Fe/H]}I-{\rm [Fe/H]}I_{Carretta09}$=~-0.01 dex ($\sigma$=~0.06 dex) and 
${\rm [Fe/H]}II-{\rm [Fe/H]}II_{Carretta09}$=~+0.13 dex ($\sigma$=~0.06 dex), for neutral and ionized 
iron abundances, respectively.

The two clusters have very similar iron content, with a difference ${\rm [Fe/H]}I_{NGC~5694}-{\rm [Fe/H]}I_{NGC6397}$=~+0.02 dex
and ${\rm [Fe/H]}II_{NGC~5694}-{\rm [Fe/H]}II_{NGC6397}$=+0.07 dex, hence we can safely assume that they have indeed the same metallicity. 


\section{The chemical composition of NGC~5694}

\subsection{Chemical homogeneity and iron abundance}
\label{homo}

As a first step in investigating the chemical properties of the cluster we check if the observed distributions 
of abundance ratios are consistent with chemical homogeneity, i.e. if all the stars have the same abundance of 
a given element within the uncertainties or if there is an {\em intrinsic} spread. To this aim we use the Maximum 
Likelihood (ML) algorithm described in \citet{m12}, that provides the mean, the intrinsic spread ($\sigma_{int}$) 
and the respective uncertainties for a given set of abundances, properly taking into account individual observational 
errors, in the hypothesis that the considered distribution can be approximated by a Gaussian.

In Table~4 we summarize the mean value and the intrinsic spread of each abundance ratio, with the respective 
uncertainties as derived with the ML algorithm. 
Excluding Na and Al (discussed separately in Sect.~\ref{nao}), the target stars do not show significant spread 
in any of the derived abundance ratios. For [Ni/Fe] we detected a formal non-null spread but fully consistent with zero, 
within the errors ($\sigma_{[Ni/Fe]}$=+0.02$\pm$0.03). 

While the considered sample is too small to draw a firm conclusion concerning the cluster as a whole, these results 
strongly suggest that the stars in NGC~5694 are as chemical homogeneous as those of typical GCs \citep{carretta09c}. 
For this reason, in the following we will adopt the {\em average} cluster abundance, as reported in Table~4, instead 
of the abundances of individual stars for all the elements except Na and Al. This will make easier the comparison with other GCs.

For La and Eu we have only upper limits to the abundance, hence we do not provide intrinsic spreads and we assume 
as typical upper limits for the cluster the values of the upper limits derived for the star with the highest SNR.

{\em The average iron content derived from the six giants analysed in this work is 
[Fe/H]I=--1.98$\pm$0.03 from neutral iron lines and [Fe/H]II=--1.83$\pm$0.01 from single ionized lines.}
As discussed in Sect.~\ref{nlte}, the offset between the two iron abundances is likely due to over-ionization 
effects that can over-estimate the abundances derived from neutral lines. Thus, we rely on the [Fe/H] ratio from 
single ionized lines as reliable estimate of the cluster metallicity. 

\begin{figure}
\includegraphics[width=84mm]{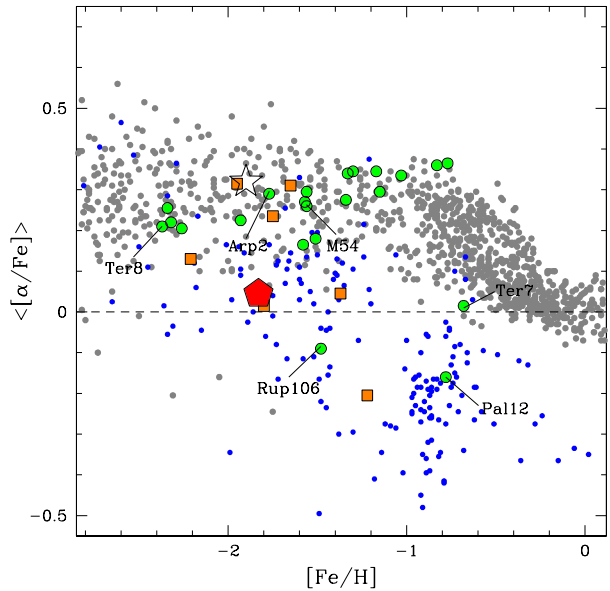}
\caption{$\langle$[$\alpha$/Fe]$\rangle$=[0.5(Ca+Ti)/Fe] as a function of [Fe/H] for different samples of stars and globular clusters. 
NGC~5694 is represented as a red filled pentagon, while the empty starred symbol indicates NGC~6397
(from our homogeneous analysis, see Section \ref{6397}). 
Filled green circles are Galactic GCs; filled orange squares are old GCs of the LMC.
Small grey filled circles are Galactic field stars; small blue circles are field stars in dSph satellites of the Milky Way. 
The dashed line marks the solar value of $\langle$[$\alpha$/Fe]$\rangle$.}
\label{alf}
\end{figure}

\subsection{Comparison with other stellar systems}

In Fig.2--5 we compare the average abundance of NGC~5694 as derived in the present study (large filled pentagon) with several sets of stars and star clusters, for a few key abundance ratios.
Our benchmark comparison cluster NGC~6397 (see Sect.~\ref{6397}, above) is represented as a large empty star.

The other data sources and symbols are the following:

\begin{itemize}

\item Galactic field stars (grey points) 
are from \citet{edvardsson93}, \citet{burris00}, \citet{fulbright00}, \citet{stephens02}, \citet{mishenina02},  
\citet{gratton03}, \citet{reddy03}, \citet{bihain04}, \citet{barklem05}, \citet{bensby05}, and \citet{reddy06}.

\item Stars from dwarf spheroidal (dSph) galaxies (blue points) are from 
\citet[][Carina, Fornax, Leo~II, Sculptor, Ursa Minor and Sextans]{shetrone01,shetrone03}, 
\citet[][Sagittarius]{sbordone07},  
\citet[][Carina]{koch08}, 
\citet[][Draco]{cohen09},
\citet[][Fornax]{letarte10},
\citet[][Ursa Maior~II and Coma Berenices]{frebel10},
\citet[][Carina]{venn12} and  
\citet[][Carina]{lemasle12}.

\item Data for Galactic GCs (large green filled circles) for $\alpha$-elements are from  \citet{carretta09}, for Cu from \citet{simmerer03}, 
for Ba from \citet{dorazi}, for Eu from \citet{sneden97,sneden04}, \citet{ivans99,ivans01},, \citet{ramirez02}, \citet{lee02}, \citet{james04}.
GCs associated with the Sagittarius (Sgr) dSphs or with its tidal stream are labeled with their names: 
M~54 \citep{brown99,carretta_m54}, Arp 8 \citep{mottini}, Palomar 12 \citep{cohen04}, Terzan 7 
\citep{sbordone07} and Terzan 8 \citep{mottini}. We did the same also for Ruprecht~106, strongly suspected as being of 
extra-galactic origin because of its lower-than-average age \citep{dott_106} and peculiar abundance pattern 
\citep{brown97}\footnote{Rup~106 was also tentatively associated with the so-called ''Orphan Stream'' by \citet{fell_106},
 but \citet{heidi_orphan} rejected the association based on a more detailed determination of the orbit of the stream.}.

\item  Old (age $\ga$ 10 Gyr) GCs in the Large Magellanic Cloud (LMC) are represented as filled orange squares. The data are from \citet{johnson06} 
and \citet{m10}.

\end{itemize}

\begin{figure}
\includegraphics[width=84mm]{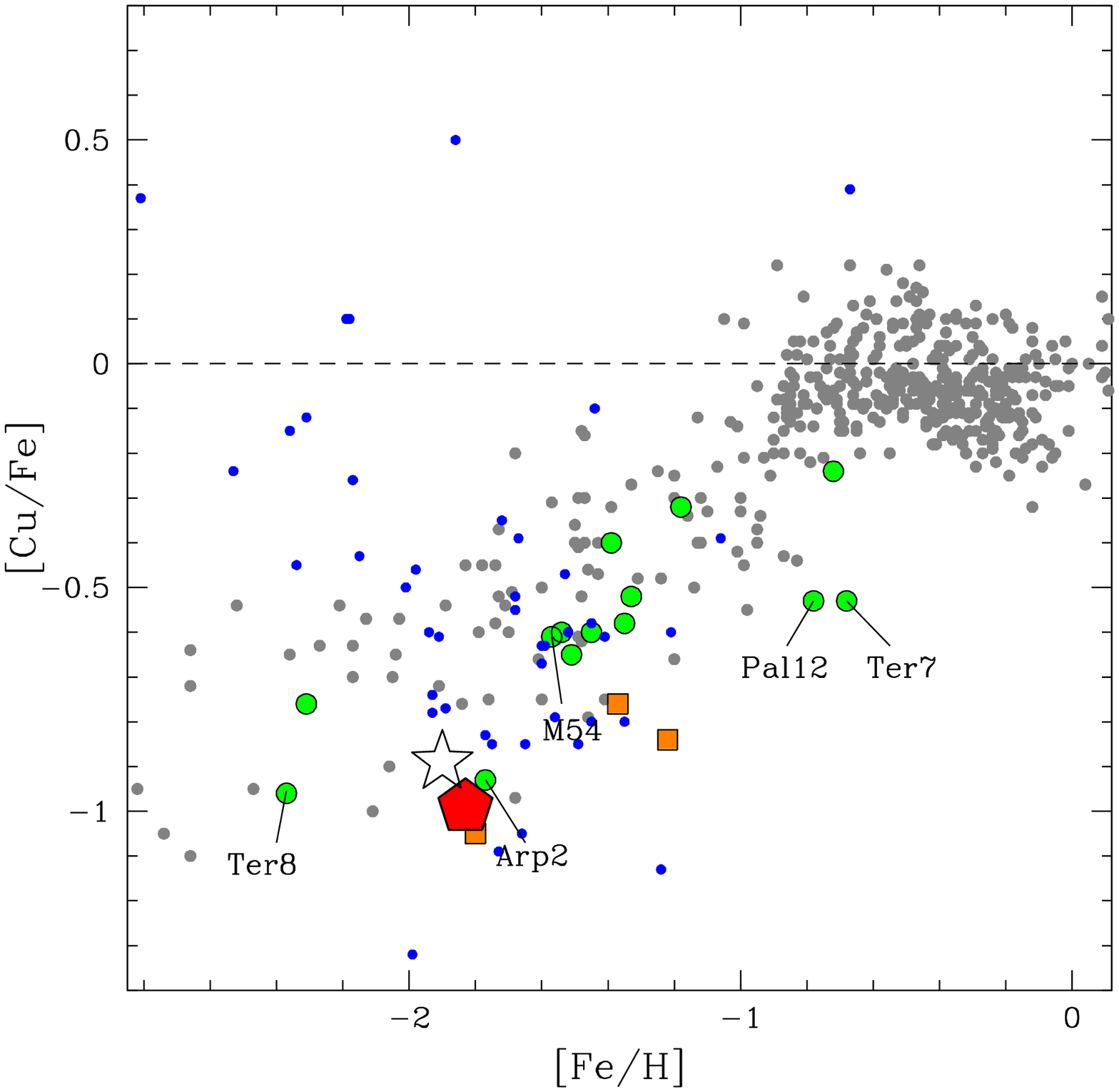}
\caption{Behaviour of [Cu/Fe] as a function of [Fe/H]. Same symbols of Fig.~\ref{alf}.
}
\label{cop}
\end{figure}

\subsubsection{Average abundances: $\alpha$ elements}

The [$\alpha$/Fe] ratios of NGC~5694 (see Tables 3 and 4) are around the solar value for O, Mg, Ca, and Ti, while only for Si the cluster shows an enhanced abundance ratio, 
more typical of Halo stars and clusters ([Si/Fe]=+0.30$\pm$0.03). In Fig.~2 we show the position of NGC~5694 in the 
$\langle$[$\alpha$/Fe]$\rangle$=[0.5(Ca+Ti)/Fe] versus [Fe/H] plane. The adopted combination of $\alpha$ elements is intended to be broadly representative 
of the group while (a) guaranteeing the largest sample of stars and GCs to compare with, since the abundances of these two elements are relatively 
easy to measure, and (b) being limited to elements that do not show significant abundance spread in ordinary GCs due to self-enrichment processes 
\citep[like O and Mg;][]{gratton}. 

It is quite clear that the [$\alpha$/Fe] ratio of NGC~5694 is significantly lower than any MW or LMC
cluster having [Fe/H]$\le -1.5$; moreover, it lies at the lower limit of the distribution of the bulk of Halo stars of similar metallicity. 
The direct ~-- and strictly differential ~-- comparison with NGC~6397 shows that, while the abundance of Si in the two clusters is not too different, 
Ca and Ti abundances in NGC~5694 are much (and very significantly) lower than in NGC~6397. O and Mg were not measured in our homogeneous analysis of 
this cluster (Sect.~\ref{6397}), but were obtained by \citet{carretta09}. They found significant intrinsic spread in the abundance of both elements and 
mean values of [O/Fe]=+0.29 and [Mg/Fe]=+0.46, both significantly larger than what we get for NGC~5694.
Hence, we must conclude that  the difference in the abundance pattern of $\alpha$ elements between NGC~5694 and classical Halo GCs is large, strongly 
significant and robustly established.

In Fig.~2, NGC~5694 appears to fall straight on the sequence defined by dSph stars, that is also populated by some Sgr 
and LMC clusters and by Rup~106. This clearly suggests  an origin in the same kind of environment ~-- a dwarf galaxy ~-- also for NGC~5694.

\begin{figure}
\includegraphics[width=84mm]{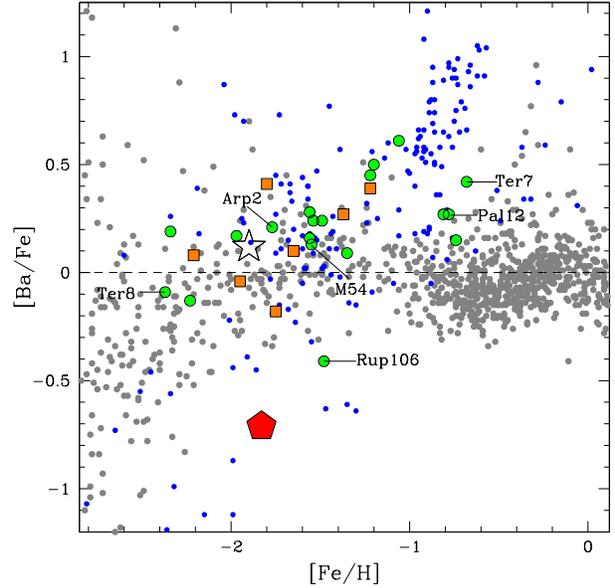}
\caption{Behaviour of [Ba/Fe] as a function of [Fe/H]. Same symbols of Fig.~\ref{alf}.
}
\label{bar}
\end{figure}

\begin{figure}
\includegraphics[width=84mm]{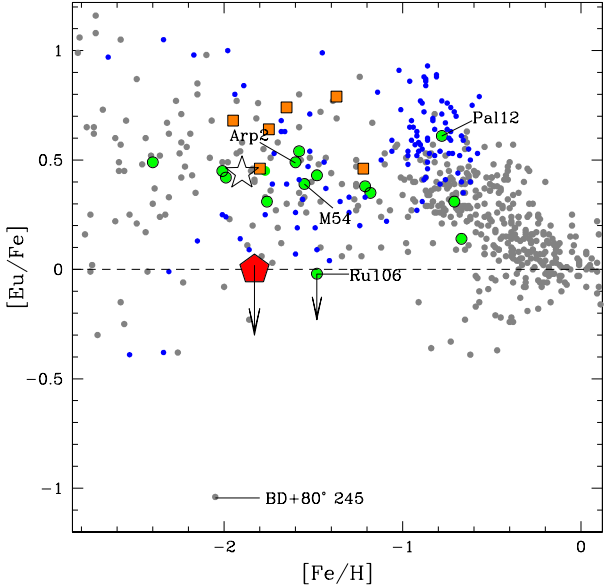}
\caption{Behaviour of [Eu/Fe] as a function of [Fe/H]. Same symbols of Fig.~\ref{alf}. 
Additionally, the position of the Galactic star BD+$^{\circ}$~245 is shown \citep{carney97}.
}
\label{eur}
\end{figure}

\subsubsection{Average abundances: iron-peak elements, copper, r- and s-elements}

Iron-peak elements like Sc, V, Mn, Co and Cr have sub-solar abundance ratios, ranging 
from a mild depletion with respect to the solar value ([Cr/Fe]=~--0.14$\pm$0.02 dex) to 
a significant depletion (as observed for Mn, with [Mn/Fe]=~--0.48$\pm$0.07 dex). 
Among the iron-peak elements, only Ni has a solar-scaled abundance ratio ([Ni/Fe]=--0.06$\pm$0.01 dex). 
Differences between NGC~5694 and NGC~6397 of the order of $\sim$0.25 dex (see Table 4) have been 
detected for [Sc/Fe], [V/Fe] and [Co/Fe], while the [Mn/Fe], [Cr/Fe]  and [Ni/Fe] abundance ratios are compatible 
between the two clusters (with differences smaller than 0.1 dex).


NGC~5694 is remarkably deficient in Cu, [Cu/Fe]=--0.99$\pm$0.06, as noted also by L06.
While the abundance of copper is at the lower limit of the Galactic halo stars distribution, it is 
pretty similar to that of NGC~6397, as well as to other metal-poor GCs (see Fig.~3).

An overall depletion of the slow neutron-capture elements is found, with [Y/Fe]II=--0.77$\pm$0.03 and [Ba/Fe]II=--0.71$\pm$0.06. 
The upper limits derived for [La/Fe]II and [Eu/Fe]II indicate sub-solar abundance ratios for both these elements. 
{\em The abundance ratios of Y, Ba, La, and Eu in NGC~5694 are significantly lower than in NGC~6397}. Fig.~4 shows at a first 
glance that the cluster has [Ba/Fe] much lower than any other considered GC and than any Galactic field stars in the considered 
sample having [Fe/H]$\ga$-2.2 (except one, see below)\footnote{See \citet{carney97} for a small sample of Ba deficient Galactic 
stars in the solar neighborhood.}. Only a couple of stars from the Carina dSph populates the same region of the diagram as NGC~5694, 
while the only other cluster showing a [Ba/Fe] ratio significantly lower than other GCs of similar metallicity is Rup~106.

The similarity between NGC~5694 and Rup~106 emerges also in the [Eu/Fe] vs. [Fe/H] plane, shown in Fig.~5. Both clusters stand 
out as remarkably Eu-deficient with respect to other GCs and both 
Galactic and dSph stars. In our ''comparison sample'' there is only one star that is strongly deficient in Eu, namely
the solar-neighborhood sub-giant 
star BD+80$^{\circ}$~245, discovered by \citet{carney97}, with [Fe/H]=--2.05 and and [Eu/Fe]=--1.04 \citep{fulbright00}. 
The star displays also sub-solar average [$\alpha$/Fe] ratio and [Ba/Fe]=--1.87; given its peculiar abundance pattern and 
large  apogalactic distance it has been indicated as having an extra-Galactic origin \citep{carney97}.


\begin{figure}
\includegraphics[width=84mm]{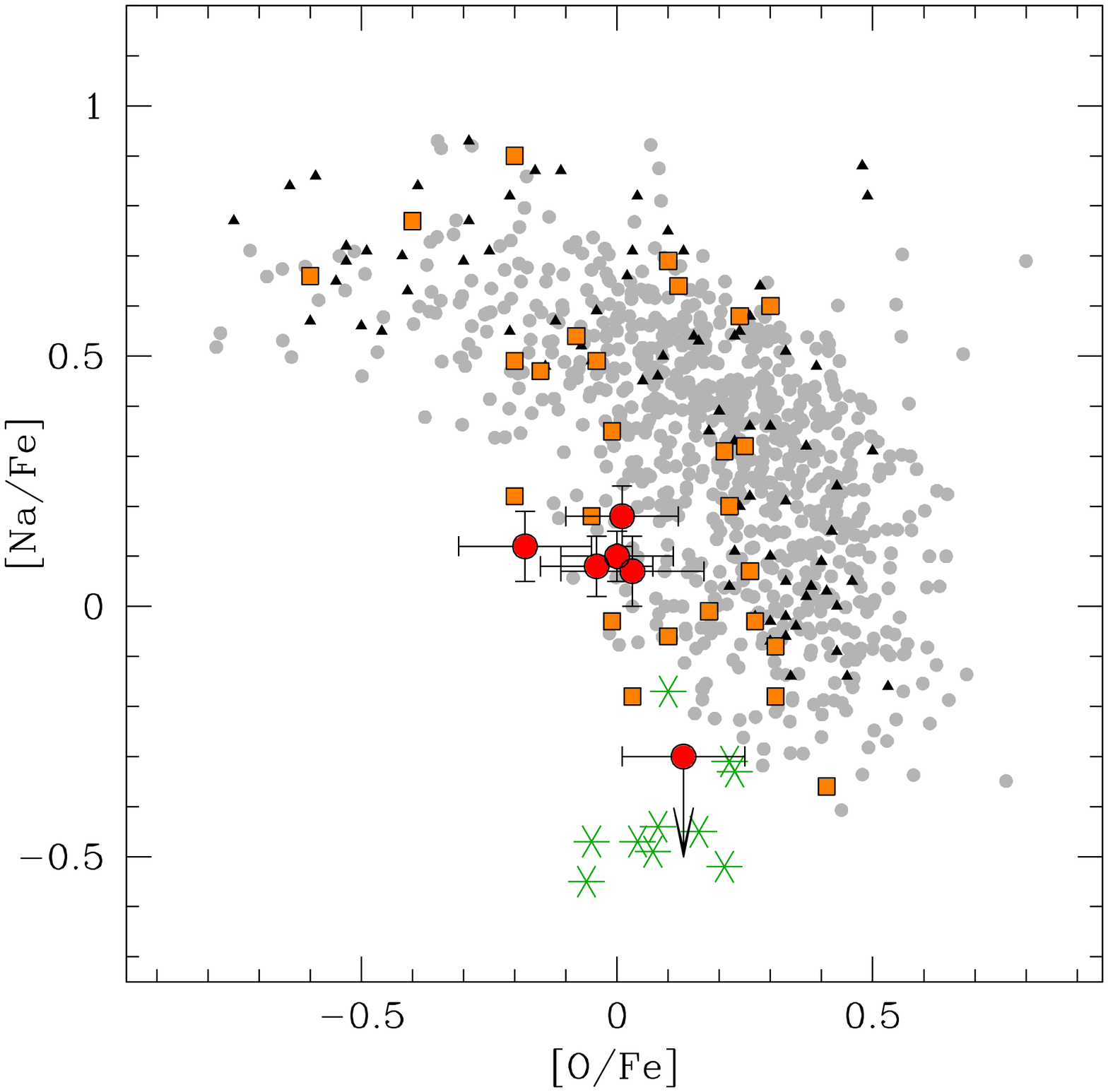}
\caption{Behaviour of [Na/Fe] as a function of [O/Fe]. Red circles are the individual 
stars of NGC~5694, in comparison with the individual stars in the Galactic GC sample 
by \citet[][grey circles]{carretta}, in M54 \citep[][black triangles]{carretta_m54}, 
in the LMC old GCs \citep[][orange squares]{johnson06,m10} and in the  GCs of extra-Galactic origin: 
Ruprecht 106, Palomar 12 and Terzan 7 \citep[][green asterisks]{brown97,cohen04,sbordone07}.
}
\label{naof}
\end{figure}

\begin{figure}
\includegraphics[width=84mm]{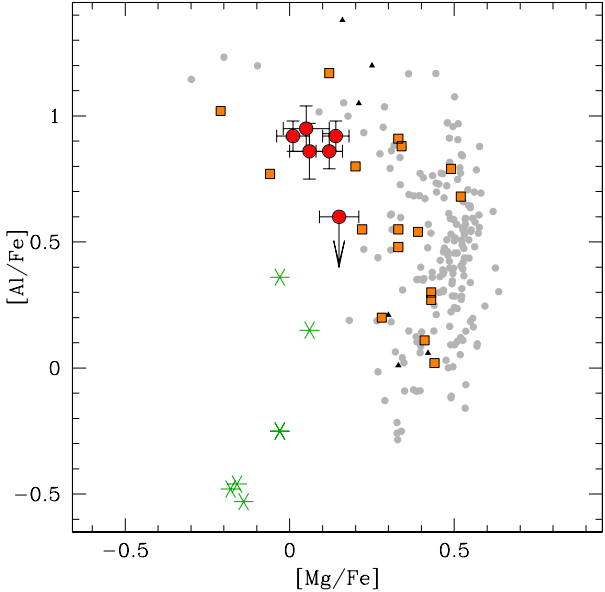}
\caption{Behaviour of [Mg/Fe] as a function of [Al/Fe]. Same symbols of Fig.~\ref{mgal}.}
\label{mgal}
\end{figure}

\subsection{Elements displaying abundance spreads: Na, O, Mg, Al}
\label{nao}

An impressive amount of photometric and spectroscopic information collected in the past decade
has reversed the classical paradigm of  GCs as simple stellar populations, revealing the co-existence 
in the same cluster of different stellar generations with very small (less than $\sim$200 Myr) age differences 
but distinct chemical abundance patterns \citep[see][and references therein]{carretta,gratton}. 
As discussed by many authors \citep[see e.g.][and references therein]{val11},
it is generally believed that the chemical enrichment of the GCs should have been driven by polluters 
like, e.g. AGB stars and/or fast rotating massive stars,  ejecting into the intra-cluster medium gas processed 
by CNO, Ne-Na and Mg-Al chains, leading to variations in light elements 
as He, C, N, Na, O, Mg and Al, without enrichment in iron and other heavy elements (driven by Supernovae explosions, whose ejecta 
cannot be retained in small mass systems as GCs).
The Na-O (and, to a lesser extent, Mg-Al) anti-correlation is considered as a typical feature of all the old GCs massive 
enough to retain the gas of these low-energy polluters in their early evolution and regardless of the parent galaxy \citep[see][]{carretta09,m10}.

Figs.~\ref{naof} and \ref{mgal} show the position of the individual stars of NGC~5694 (red circles) 
in the [Na/Fe]--[O/Fe] and [Mg/Fe]--[Al/Fe] planes, respectively. Five out of six stars have the same abundances of these elements, within the 
uncertainties. On the other hand, star \#95 has O and Mg abundances consistent with the others, but it shows {\em significantly lower} Al and Na 
abundances. 

As a sanity check, we analyzed also the Na resonance D doublet at 5890-5896 $\mathring{A}$, 
a feature that is very strong even in our low-metallicity stars ($\sim$250-300 m$\mathring{A}$) and is not contaminated by interstellar Na lines, 
given the systemic velocity of the cluster.
The abundance  derived from these lines is very sensitive to temperature, gravity and EW uncertainty, with respect to the other Na lines used 
in this work. Still, it provides an independent test on the difference in Na abundance between star \#95 and the other targets.
The EWs have been obtained with the IRAF task {\sl splot}, adopting 
a Voigt profile in order to properly take into account the contribution of the prominent wings. 
From NaD lines we confirm that the sodium abundance of star \#95 is indeed 
significantly lower (by $\ga$0.4 dex) with respect to the other stars.

These findings provide a first indication that the self-enrichment processes that seem common to all massive GCs occurred also in NGC~5694.
In light of the evidence collected for the Galactic GCs, we can consider star \#95 as belonging 
to the first stellar generation of the cluster, while the other five should belong to a subsequent 
generation, formed from the ejecta of the first stars.
Although the sample is too small to draw any firm conclusion on this issue, the relative fraction of 
stars of the cluster labelled as second generation stars resembles that observed in Milky Way GCs, 
where the second generation stars currently account for  $\sim$70\% of the total stellar content of the clusters \citep{carretta}.

Concerning the origin of NGC~5694, it is interesting to note that the stars considered in the present study lie at the 
margins of the [Na/Fe]--[O/Fe] and [Mg/Fe]--[Al/Fe] distribution of stars in classical GCs, and star \#95 has [Al/Fe] and, 
in particular, [Na/Fe] ratios compatible with the stars from metal-rich Sgr clusters and Rup~106 (green asterisks in Fig.~6 and~7) that, 
in turn, are clearly different from genuine Galactic GCs also in this respect 
\citep[i.e. they can be chemically tagged as extra-galactic also based on their abundances of Na, O, Mg, Al; see][]{mic_pechino}.

\section{A brief chemical history of NC5694}

The chemical composition of NGC~5694 is remarkably different with respect to the Galactic stars 
of similar metallicity. In particular, NGC~5694 exhibits at least two main chemical peculiarities 
when it is compared with Milky Way (field/GC) stars:
\begin{itemize}
\item~solar-scaled [$\alpha$/Fe] ratios, with the only exception of [Si/Fe] (that appears to be 
enhanced with respect to the solar value); 
\item deficiency of neutron-capture elements, both 
those usually associated to slow neutron-capture (as Y, La and Ba) and to rapid neutron-capture (Eu).
\end{itemize}

Since $\alpha$-elements are produced mainly in the massive stars through hydrostatic (for O and Mg) and explosive 
(for Si, Ca, and Ti) nucleosynthesis by Type II SNe, while iron is produced also by Type Ia SNe, 
the [$\alpha$/Fe] ratio is commonly adopted as tracer of the relative contribution of Type II and Ia SNe. 
In the Galactic Halo the [$\alpha$/Fe] ratio remains nearly constant around [$\alpha$/Fe]$\simeq +0.3$  up to [Fe/H]$\sim -1.0$, 
where it begins to decline  (the so-called 
{\sl knee}), reaching [$\alpha$/Fe]$\sim 0.0$ at [Fe/H]$\sim 0.0$. 
Generally, the iron content of the {\sl knee} marks the metallicity reached by the system when the ejecta 
of Type Ia SNe are well mixed in the interstellar medium and their chemical signatures dominate the gas, because the 
Type Ia SNe have longer timescales than Type II SNe.
Systems characterized by star formation rates lower than that of the Milky Way will have the {\sl knee} shifted toward 
lower metallicity, as observed in dSph galaxies \citep[see Fig.~\ref{alf} and][for a discussion and references]{tolstoy}.
Thus, the low (solar) [$\alpha$/Fe] abundance ratios (lower than those observed in the Galactic Halo 
stars of similar metallicity) clearly suggest that this cluster formed from a gas enriched by both Type II and Type Ia SNe 
and likely in an environment characterized by a star formation rate slower than that typical of the Galactic Halo.

The interpretation of the neutron-capture elements abundances is complicated by the multiplicity of 
their nucleosynthesis sites. 
Basically, the main contributors to slow neutron-capture elements (like Y, La and 
Ba) at the solar metallicity are the AGB stars with initial mass between $\sim$1 and $\sim$4 $M_{\odot}$ 
during the thermal pulses stage \citep[see e.g.][]{busso}, the so-called main s-component. 
Slow neutron-capture processes occur also in massive (M$>$8$M_{\odot}$) stars 
during the convective core He burning phase, the so-called weak s-component 
\citep[see e.g.][]{raiteri,pignatari}.
These elements are not totally built from s-processes because a small fraction (of the order of $\sim$20\% for 
these elements) is produced from rapid neutron-captures. The efficiency of the r-process can be estimated 
from the analysis of Eu, considered a pure r-process element, being 
formed in the solar system almost totally from r-process nucleosynthesis \citep[$\sim$97\%, see e.g.][]{burris00}.
The precise site of the r-process is still debated but it is usually associated to massive 
stars where a relevant neutron flux is available. 
The chemical abundances of neutron-capture elements in NGC~5694 are 
unusual for both field and GC stars in the Galactic Halo and clearly point out 
that NGC~5694 has formed from gas enriched in a different way with respect to the Halo.

In the Milky Way, no GCs with such a low [Ba/Fe] ratio is observed \citep[see][]{dorazi}, while among field 
stars [Ba/Fe] decreases in the metal-poor regime, at [Fe/H]$<$--2.5 dex \citep{barklem05}. 
Also, NGC~5694 is remarkably deficient in Eu ([Eu/Fe]$<$0.0), while the stars of similar metallicity 
observed in the Milky Way (as well as in the LMC and dSphs) are characterized by enhanced values of
[Eu/Fe]. 
The Eu depletion seems to suggest that NGC~5694 formed from gas where the enrichment by 
r-process has been less efficient than in other galaxies of the Local Group studied so far. 
This finding could explain (at least partially) the low Y, Ba and La abundances. Because 
the contribution by AGB stars starts to be relevant after $\sim$100-300 Myr from the first burst of 
star formation in the galaxy where the cluster formed, these elements are produced mainly from 
r-processes.

L06 include among the chemical peculiarities of NGC~5694 also the low [Cu/Fe] ratio. 
In light of our differential analysis with NGC~6397, that shows a [Cu/Fe] ratio higher than that of 
NGC~5694 by only 0.1 dex, we are cautious to over-interpret the Cu abundance in the cluster.
The stars in the Milky Way show a relevant scatter of [Cu/Fe] in the metal-poor regime: 
the bulk of the stars around [Fe/H]$\sim$--2 dex has [Cu/Fe]$\sim$--0.7 dex, but both the sample 
of \citet{mishenina02} and \citet{bihain04} include stars with [Cu/Fe]$\sim$--1 dex at the metallicity of 
NGC~5694. On the other hand, the sample of GCs by \citet{simmerer03} shows higher (by about 0.3 dex) values of [Cu/Fe] for
metal-poor clusters. 
Unfortunately, the dataset by \citet{simmerer03} does not include 
NGC~6397 and we cannot perform a direct comparison between the two analyses
\footnote{\citet{simmerer03} derived the Cu abundance for the majority of their stars from the 
Cu line at 5782 $\mathring{A}$ and only for a sub-sample of star they are able to measure the line 
at 5105 $\mathring{A}$, used in our work. In order to check for possible offset between the two transitions, 
we measured them in the solar flux spectrum of \citet{neckel}, finding differences in the derived abundances 
less than 0.05 dex (and in excellent agreement with the solar Cu abundance).                
Also, we checked that there are no difference between the employed oscillator strengths, the hyperfine components and 
the isotopic ratio for the 5105 $\mathring{A}$ between us and \citet{simmerer03}.}.

\section{Who are the relatives of NGC~5694?}

Our chemical analysis confirms and extends the first findings by L06, definitely tagging NGC~5694 
as a {\sl foster son} of the Galactic Halo, therefore a cluster of extra-Galactic origin.
Its abundance pattern is significantly different from the other Galactic GCs of similar metallicity, as well as from the bulk of metal-poor 
Halo field stars. The cluster most likely originated in a dwarf galaxy having a slower chemical evolution and 
a lower enrichment by the r-process  with respect to the Milky Way \citep[see][]{tolstoy}.

NGC~5694 is, by far, the most metal-poor GC chemically tagged as an interloper of the Galactic halo.
We note that clusters having similar metallicity, that are known to be members of the Sgr dSph (Ter~8, Arp~2, M~54) do not show 
the chemical anomalies that allowed L06 and us to recognize NGC~5694 as an incomer: they have $\alpha$, Eu and Ba 
abundances fully compatible with the classical Halo population \citep[see also][]{mottini2}. Hence, an association of NGC~5694 with the Sgr 
galaxy or with its tidal stream can be excluded on the basis of the chemical abundance pattern alone \citep[thus confirming the conclusions by][]{lm10}.

None of the dSphs studied so far fully matches the chemical composition of NGC~5694 at the same metallicity level, since, in particular, 
they display higher abundances of neutron-capture elements, on average. 
Indeed, the dSph stars with an iron content around [Fe/H]$\sim$--2 cover a large range of [Ba/Fe] values
typically with [Ba/Fe]$>$--0.5 dex (but basically compatible with the Galactic stars), and {\em enhanced} [Eu/Fe] 
\citep[see for instance Fig.~13 in][]{tolstoy}. Still, the fraction of stars reaching [Eu/Fe]$\la +0.2$ is larger among 
dSphs than among the Halo population, in the considered metallicity regime.
Two Ba-poor stars have been identified in the Carina dSph by \citet[][namely MKV0397]{lemasle12} 
and \citet[][namely \#5070]{venn12}. MKV0397 has [Fe/H]I=--1.99 and [Ba/Fe]II=--0.87, but with [Ca/Fe]=+0.39 (unfortunately, the Eu abundance is not provided 
for this star). \#5070 has [Fe/H]I=--2.15 and [Ba/Fe]II=--1.12, showing also solar abundances for [Ca/Fe] and [Ti/Fe] but a strong (--0.36) depletion for 
[Mg/Fe], while a slightly over-solar (+0.13) [Eu/Fe] ratio is found.
A Bar-poor star has been found also in the Draco dSph by \citet[][namely XI-2]{cohen04}, having [Fe/H]I=--1.99 and [Ba/Fe]II=--1.12, but 
[Ca/Fe] and [Ti/Fe] are heavily depleted ($<$--0.3 dex).
We conclude that the available data are not incompatible with the hypothesis that NGC~5694 was born in a system (chemically) 
similar to present-day dSphs, but the chemical properties of the cluster are pretty extreme with respect to the overall distribution of dSph 
stars. 

Also Ultra Faint Dwarf galaxies (UFD) typically have stars as metal-poor as NGC~5694 with solar or sub-solar [$\alpha$/Fe] ratios \citep{ufd}. 
While abundance estimates for r-process elements are largely lacking for these systems, it is very intriguing to note that stars with 
[Fe/H]$\simeq -2.0$,  [$\alpha$/Fe]$\simeq 0.0$ {\em and} [Ba/Fe]$\la -0.5$, i.e. fully analogous to  NGC~5694 stars, have been actually 
observed in some UFD \citep[namely CVn~II and Leo~IV, see][]{franc_ufd}.

In a similar way, we check also the possibility that NGC~5694 can be associated to the LMC. 
\citet{johnson06} and \citet{m10} provide conflicting results about the [$\alpha$/Fe] ratios in 
old LMC GCs. The clusters analysed by \citet{johnson06} have solar-scaled [$\alpha$/Fe], compatible 
with the dSph stars and with the abundances in NGC~5694. On the hand, the GCs discussed by 
\citet{m10} have enhanced [$\alpha$/Fe] ratios, compatible with those of the Halo stars. However, both 
the analyses point to solar or enhanced [Ba/Fe] ratios and enhanced [Eu/Fe] for the LMC GCs. 
Thus, the different abundances of the neutron-capture elements (and in particular the different 
level of enrichment by the r-process) do not support the idea that NGC~5694 formed in the LMC.

 It is interesting to note that the chemical oddities that make NGC~5694 such a peculiar object are 
 ~-- to some extent ~-- shared by Rup~106, a cluster whose extra-Galactic origin has been suspected since long time \citep{r106,lin,ffp95}. 
As in NGC~5694, the two giant stars of Rup~106 discussed by \citet{brown97}  
have [$\alpha$/Fe], [Ba/Fe] and [Eu/Fe] ratios lower than those of the Halo field stars of similar metallicity. The two clusters lie in the 
same broad region of the outer halo, having Galactic longitude, latitude and galactocentric distance 
(l,b,R$_{GC})_{NGC~5694}=(331.1\degr, +30.4\degr, 29.4$~kpc) and (l,b,$R_{GC})_{Rup106}=(300.9\degr, +11.7\degr, 18.5$~kpc), according to 
\citet[][2010 edition]{harris}. They also have very similar galactocentric radial velocity \citep[$V_{GSR}= -233$ and $-231$ km/s, for Rup~106 and NGC~5694, respectively;][]{lm10}. 
By taking into account its chemical composition, as derived in the present study,
we confirm that NGC~5694 is as old as the oldest Galactic GC \citep[in particular, as old or even slightly older than NGC~6397;][]{angeli}.
This implies that Rup106 is a couple of Gyr younger than NGC~5694 \citep[see][]{dott10,dott_106}. In turn, 
since Rup106 is also $\sim 0.5$~dex more metal-rich than NGC~5694, this may hint at a common age-metallicity relation, 
as observed in the GC systems of some dSphs \citep[see, e.g.][]{ter8,forbes}.
Given all the above considerations, the hypothesis of a common origin for the two clusters, in the same (now disrupted) 
dwarf satellite of the Milky Way, cannot be easily dismissed.

\section*{Acknowledgments}

The authors warmly thank the referee for his/her valuable comments and suggestions.
M.B. acknowledges the financial support from PRIN MIUR, project: {\em The
Chemical and Dynamical Evolution of the Milky Way and Local Group Galaxies},
prot. 2010LY5N2T. 
M.C. and P.A. are supported by the Chilean Ministry for the Economy, Development, and Tourism's Programa 
Iniciativa Cient\'{i}fica Milenio through grant P07-021-F, awarded to The Milky Way Millennium Nucleus; 
by the BASAL Center for Astrophysics and Associated Technologies (PFB-06); by Proyecto Fondecyt Regular \#1110326; and by Proyecto 
Anillo ACT-86. 
We thank E. Carretta for sharing with us the unpublished data 
for the Ti abundances in Galactic GCs.

\begin{table*}
\begin{minipage}{120mm}
\caption{Fundamental parameters for target stars.}
\begin{tabular}{lccccccccc}
\hline
ID &  G95-ID &  RA$_{J2000}$ &  Dec$_{J2000}$ & B & V &  RV & $T_{eff}$ & logg  & $v_t$ \\
   &          &  [deg]  &  [deg]    &   &   &  [km/s]    & [K]         &       & [km/s] \\
\hline
   37 & 62   &   219.8853105  &  -26.5527808  &  17.025  &    15.644   & --138.6$\pm$0.3  & 4180  & 0.65	 &   1.90  \\
   57 & 51   &   219.8833781  &  -26.5442684  &  17.180  &    15.924   & --140.0$\pm$0.4  & 4324  & 0.86	 &   1.80  \\
   68 &110   &   219.9161622  &  -26.5452373  &  17.276  &    16.094   & --138.5$\pm$0.3  & 4414  & 0.99	 &   1.80  \\
   82 &109   &   219.9107318  &  -26.5464133  &  17.416  &    16.292   & --151.2$\pm$0.4  & 4488  & 1.11	 &   1.80  \\
   95 & 28   &   219.8684541  &  -26.5313730  &  17.498  &    16.414   & --137.0$\pm$0.4  & 4540  & 1.19	 &   1.80  \\
  105 & 68   &   219.8854188  &  -26.5660950  &  17.560  &    16.518   & --136.9$\pm$0.5  & 4596  & 1.26	 &   1.60  \\
\hline
\end{tabular}
\end{minipage}
\end{table*}

\begin{table*}
\begin{minipage}{150mm}
\caption{Atomic data and measured equivalent widths of the UVES targets. 
The complete table is available in the electronic form.}
\begin{tabular}{lcccccc}
\hline
{\rm Star} & $\lambda$ & Ion  & loggf  &  $\chi$   &  EW  & $\sigma_{{\rm EW}}$  \\
          &       ($\mathring{A}$) &        &    & (eV)    & (m$\mathring{A}$)   & (m$\mathring{A}$)  \\
\hline
NGC~5694-37 &   4882.1  & {\rm Fe~I}	 &	 -1.640        &      3.420   &    48.00     &       5.69         \\ 
NGC~5694-37 &   4927.4  & {\rm Fe~I}	 &	 -2.070        &      3.570   &    20.20     &       3.02         \\ 
NGC~5694-37 &   4950.1  & {\rm Fe~I}	 &	 -1.670        &      3.420   &    53.50     &       4.75      \\ 
NGC~5694-37 &   4985.2  & {\rm Fe~I}	 &	 -0.560        &      3.930   &    76.90     &       5.29       \\ 
NGC~5694-37 &   5002.7  & {\rm Fe~I}	 &	 -1.530        &      3.400   &    57.60     &       3.61         \\ 
NGC~5694-37 &   5014.9  & {\rm Fe~I}	 &	 -0.300        &      3.940   &    79.60     &       3.17         \\ 
NGC~5694-37 &   5028.1  & {\rm Fe~I}	 &	 -1.120        &      3.570   &    79.20     &       5.20         \\ 
NGC~5694-37 &   5044.2  & {\rm Fe~I}	 &	 -2.020        &      2.850   &    60.70     &       4.70         \\ 
NGC~5694-37 &   5090.7  & {\rm Fe~I}	 &	 -0.440        &      4.260   &    60.10     &       3.74         \\ 
\hline
\end{tabular}
\end{minipage}
\end{table*}

\begin{center}
\begin{table*}
\begin{minipage}{120mm}
\caption{Abundance ratios for individual target stars.}
\begin{tabular}{lcccccc}
\hline
ID &  [Fe/H]~I &  [Fe/H]~II &  [O/Fe] & [Na/Fe] & [Mg/Fe]  \\
\hline
Sun  &  7.50   &  7.50     &    8.76    &  6.33  & 7.58  \\
\hline
 37 &  -2.01$\pm$0.07  &   -1.83$\pm$0.06 &   0.01$\pm$0.11 (0.09) &  0.18$\pm$0.06 (0.08)   &  0.12$\pm$0.04  (0.06) \\
 57 &  -1.99$\pm$0.10  &   -1.89$\pm$0.05 &   0.00$\pm$0.11 (0.08) &  0.10$\pm$0.05 (0.06)   &  0.14$\pm$0.04  (0.07) \\
 68 &  -1.92$\pm$0.09  &   -1.83$\pm$0.04 & --0.18$\pm$0.13 (0.08) &  0.06$\pm$0.07 (0.07)   &  0.01$\pm$0.05  (0.07) \\
 82 &  -1.94$\pm$0.09  &   -1.78$\pm$0.05 & --0.04$\pm$0.11 (0.10) &  0.08$\pm$0.06 (0.06)   &  0.06$\pm$0.06  (0.07) \\
 95 &  -2.06$\pm$0.09  &   -1.85$\pm$0.04 &   0.13$\pm$0.12 (0.10) &  $<$--0.30              &  0.15$\pm$0.06  (0.07) \\
105 &  -1.88$\pm$0.09  &   -1.81$\pm$0.04 &   0.03$\pm$0.14 (0.10) &  0.07$\pm$0.07 (0.07)   &  0.05$\pm$0.07  (0.07) \\     
\hline
ID &    [Al/Fe]   &  [Si/Fe] &  [Sc/Fe]~II &  [Ca/Fe] &  [Ti/Fe]~I \\
\hline
Sun   & 6.47  & 7.55 &  3.17   &  6.36    &  5.02 \\
\hline
 37 &   0.86$\pm$0.07 (0.07)  &   0.36$\pm$0.07 (0.03) & --0.24$\pm$0.10  (0.07)&   0.03$\pm$0.02   (0.08) &   0.01$\pm$0.06 (0.11)\\
 57 &   0.92$\pm$0.06 (0.06)  &   0.35$\pm$0.09 (0.06) & --0.24$\pm$0.11  (0.07)&   0.05$\pm$0.02   (0.08) &   0.01$\pm$0.05 (0.13)\\
 68 &   0.92$\pm$0.06 (0.06)  &   0.21$\pm$0.07 (0.04) & --0.31$\pm$0.10  (0.07)&   0.00$\pm$0.02   (0.07) &   0.05$\pm$0.05 (0.13)\\
 82 &   0.86$\pm$0.11 (0.10)  &   0.25$\pm$0.12 (0.11) & --0.30$\pm$0.11  (0.08)&   0.02$\pm$0.03   (0.07) &   0.10$\pm$0.05 (0.12)\\
 95 &   $<$0.60 	      &   0.35$\pm$0.08 (0.07) & --0.31$\pm$0.11  (0.07)&   0.08$\pm$0.03   (0.07) &   0.14$\pm$0.06 (0.12)\\
105 &   0.95$\pm$0.09 (0.08)  &   0.30$\pm$0.09 (0.07) & --0.35$\pm$0.10  (0.06)&   0.01$\pm$0.04   (0.07) &   0.10$\pm$0.10 (0.14)\\	  
\hline
ID &  [Ti/Fe]~II & [V/Fe] &  [Cr/Fe]  &  [Mn/Fe] &  [Co/Fe] \\
\hline
Sun    &  5.02  &   4.00  &  5.39   &   4.92   &    6.25   \\
\hline
 37 &   -0.19$\pm$0.05 (0.05)  &   --0.45$\pm$0.07  (0.13)	&  -0.14$\pm$0.05  (0.11)  &  --0.60$\pm$0.17 (0.14) & --0.16$\pm$0.13  (0.08)\\
 57 &    0.00$\pm$0.09 (0.09)  &   --0.40$\pm$0.08  (0.13)	&  -0.17$\pm$0.06  (0.14)  &  --0.53$\pm$0.18 (0.16) & --0.16$\pm$0.22  (0.20)\\
 68 &   -0.14$\pm$0.05 (0.06)  &   --0.29$\pm$0.06  (0.13)	&  -0.12$\pm$0.04  (0.13)  &  --0.41$\pm$0.17 (0.14) & --0.28$\pm$0.18  (0.15)\\
 82 &   -0.13$\pm$0.07 (0.08)  &   --0.34$\pm$0.10  (0.15)	&   0.00$\pm$0.03  (0.11)  &  --0.61$\pm$0.26 (0.24) & --0.11$\pm$0.16  (0.12)\\
 95 &   -0.04$\pm$0.08 (0.08)  &   --0.35$\pm$0.12  (0.16)	&  -0.16$\pm$0.05  (0.12)  &  --0.48$\pm$0.18 (0.15) & --0.31$\pm$0.18  (0.16)\\
105 &   -0.07$\pm$0.10 (0.11)  &   --0.30$\pm$0.12  (0.16)	&  -0.13$\pm$0.08  (0.13)  &  --0.31$\pm$0.17 (0.14) & --0.12$\pm$0.20  (0.17)\\     
\hline
ID & [Ni/Fe] & [Cu/Fe] & [Y/Fe]~II &  [Ba/Fe]~II & [La/Fe]~II \\
\hline
Sun  &   6.25      &  4.21   &   2.24 &  2.13  & 1.17  \\
\hline
 37 &  -0.03$\pm$0.03 (0.08) &  --0.99$\pm$0.14 (0.09)  &   -0.64$\pm$0.12 (0.12)    &  --0.76$\pm$0.16 (0.14)   & $<$-0.25\\
 57 &  -0.10$\pm$0.03 (0.10) &  --1.03$\pm$0.15 (0.10)  &   -0.74$\pm$0.06 (0.05)    &  --0.62$\pm$0.18 (0.16)   & $<$-0.20 \\
 68 &  -0.08$\pm$0.04 (0.10) &  --0.89$\pm$0.15 (0.10)  &   -0.82$\pm$0.06 (0.06)    &  --0.69$\pm$0.16 (0.14)   & $<$-0.20 \\
 82 &  -0.05$\pm$0.04 (0.10) &  --0.88$\pm$0.15 (0.11)  &   -0.80$\pm$0.09 (0.09)    &  --0.73$\pm$0.16 (0.15)   & $<$-0.20 \\
 95 &   0.02$\pm$0.04 (0.10) &  --1.29$\pm$0.17 (0.13)  &   -0.75$\pm$0.06 (0.06)    &  --0.68$\pm$0.20 (0.18)   & $<$-0.10 \\
105 &  -0.13$\pm$0.06 (0.10) &  --1.28$\pm$0.20 (0.17)  &   -0.73$\pm$0.11 (0.11)    &  --0.76$\pm$0.17 (0.15)   & $<$0.00 \\	  
\hline
ID & [Eu/Fe]~II &  &  &  &  \\
\hline
Sun  &   0.51     &     &      &    &    \\
\hline
 37 &    $<$0.00       &    &    &   &  	  \\   
 57 &    $<$0.00       &    &    &   &  	  \\
 68 &    $<$0.10       &    &    &   &  	  \\
 82 &    $<$0.10       &    &    &   &  	  \\
 95 &    $<$0.10       &    &    &   &  	  \\
105 &    $<$0.15       &    &    &   &  	  \\ 
\hline
\end{tabular}
\end{minipage}
\end{table*}
\end{center}

\begin{table*}
\begin{minipage}{150mm}
\caption{Mean elemental abundances and intrinsic spreads for NGC~5694 and NGC~6397. 
The last column lists the difference $\delta$ between the mean abundances of NGC~5694 and NGC~6397.
Only the elements that not show intrinsic spreads are listed.}
\begin{tabular}{lccccc}
\hline
 Abundance ratio &  NGC~5694 & & NGC~6397  & \\
             & mean$\pm$error & $\sigma_{int}\pm$error & mean$\pm$error & $\sigma_{int}\pm$error  & $\delta$ \\
\hline
 {\rm [Fe/H]}    &      --1.98$\pm$0.03 &  0.00$\pm$0.04 & --2.00$\pm$0.02  &  0.00$\pm$0.02   &    +0.02   \\
 {\rm [Fe/H]II}  &	--1.83$\pm$0.01 &  0.00$\pm$0.02 & --1.90$\pm$0.01  &  0.00$\pm$0.02   &    +0.07	\\  
 {\rm [O/Fe]}    &	--0.02$\pm$0.05 &  0.00$\pm$0.05 &              --- &  ---	       &    ---	\\   
 {\rm [Mg/Fe]}   &	 +0.10$\pm$0.02 &  0.00$\pm$0.06 &             ---  & ---	       &    ---   \\ 
 {\rm [Si/Fe]}   &	 +0.30$\pm$0.03 &  0.00$\pm$0.05 &  +0.43$\pm$0.01 &   0.00$\pm$0.06   &   --0.13    \\
 {\rm [Ca/Fe]}   &	 +0.04$\pm$0.01 &  0.00$\pm$0.01 &  +0.37$\pm$0.01 &   0.00$\pm$0.01   &   --0.33    \\ 
 {\rm [Sc/Fe]II} &	 -0.29$\pm$0.04 &  0.00$\pm$0.04 & --0.05$\pm$0.02 &   0.00$\pm$0.04   &   --0.24    \\   
 {\rm [Ti/Fe]}   &	 +0.05$\pm$0.02 &  0.00$\pm$0.04 &  +0.27$\pm$0.01 &   0.00$\pm$0.03   &   --0.22    \\ 
 {\rm [Ti/Fe]II} &	--0.13$\pm$0.02 &  0.00$\pm$0.04 &  +0.21$\pm$0.01 &   0.02$\pm$0.03   &   --0.34    \\ 
 {\rm [V/Fe]}   &	--0.35$\pm$0.03 &  0.00$\pm$0.09 & --0.14$\pm$0.03 &   0.00$\pm$0.04   &   --0.21	\\   
 {\rm [Cr/Fe]}   &	--0.14$\pm$0.02 &  0.00$\pm$0.02 & --0.16$\pm$0.01 &   0.00$\pm$0.03   &    +0.02   \\ 
 {\rm [Mn/Fe]}   &	--0.48$\pm$0.07 &  0.00$\pm$0.09 & --0.55$\pm$0.04 &   0.00$\pm$0.06   &    +0.07	\\  
 {\rm [Co/Fe]}   &	--0.19$\pm$0.06 &  0.00$\pm$0.07 &   0.07$\pm$0.04 &   0.00$\pm$0.05   &   --0.26   \\   
 {\rm [Ni/Fe]}   &	--0.06$\pm$0.01 &  0.02$\pm$0.02 &   0.00$\pm$0.01 &   0.00$\pm$0.01   &   --0.06    \\  
 {\rm [Cu/Fe]}   &	--0.99$\pm$0.06 &  0.00$\pm$0.09 & --0.89$\pm$0.04 &   0.00$\pm$0.10   &   --0.10	\\   
 {\rm [Y/Fe]II}  &	--0.77$\pm$0.03 &  0.00$\pm$0.03 & --0.35$\pm$0.01 &   0.00$\pm$0.03   &   --0.42	\\ 
 {\rm [Ba/Fe]II} &	--0.71$\pm$0.06 &  0.00$\pm$0.06 &   0.12$\pm$0.02 &   0.00$\pm$0.03   &   --0.83	\\   
 {\rm [La/Fe]II} &	      $<$--0.25 &     ---	 &   0.24$\pm$0.02 &   0.00$\pm$0.04   &  	---\\   
 {\rm [Eu/Fe]II} &	      $<$0.00	&     ---	 &   0.44$\pm$0.01 &   0.00$\pm$0.03   &  	---\\   
\hline
\end{tabular}
\end{minipage}
\end{table*}

\end{document}